\documentclass[floatfix,aps,showpacs,preprint,preprintnumbers,amsmath,amssymb]{revtex4}

\usepackage{graphicx}
\usepackage{dcolumn}
\usepackage{bm}

\begin{document}

\newcommand{\MAG}{Fe$_3$O$_4\;$}
\newcommand{\MOS}{M\"{o}ssbauer\;}
\newcommand{\NP1}{1}
\newcommand{\NP2}{2}
\newcommand{\NP3}{3}
\newcommand{\NP4}{4}
\newcommand{\NP5}{5}
\newcommand{\NP6}{6}
\newcommand{\NP7}{7}

\title{Spin Disorder and Magnetic Anisotropy in $Fe_3O_4$
Nanoparticles}

\author{E. Lima Jr., A. L. Brandl, A. D. Arelaro, and G. F. Goya\footnote{Present
address: Instituto de Nanociencia de Arag\'{o}n, Universidad de Zaragoza, Zaragoza, Spain }\email }
\email{goya@unizar.es}
\affiliation{Instituto de F\'isica, Universidade de S\~ao Paulo\\
CP 66318, 05315-970, S\~ao Paulo, Brazil}

\date{\today }

\begin{abstract}
We have studied the magnetic behavior of dextran-coated magnetite
(Fe$_3$O$_4$) nanoparticles with median particle size
$\left<d\right>=8$ $nm$. Magnetization curves and in-field
M\"ossbauer spectroscopy measurements showed that the magnetic
moment $M_S$ of the particles was much smaller than the bulk
material. However, we found no evidence of magnetic irreversibility
or non-saturating behavior at high fields, usually associated to
spin canting. The values of magnetic anisotropy $K_{eff}$ from
different techniques indicate that surface or shape contributions
are negligible. It is proposed that these particles have bulk-like
ferrimagnetic structure with ordered A and B sublattices, but nearly
compensated magnetic moments. The dependence of the blocking
temperature with frequency and applied fields, $T_B(H,\omega)$,
suggests that the observed non-monotonic behavior is governed by the
strength of interparticle interactions.
\end{abstract}

\pacs{75.50.Tt, 75.50.Gg, 75.30.Gw, 76.80.+y}%
\maketitle

\section{\label{sec:1}INTRODUCTION}

Iron oxide nanoparticles have gained technological significance
after they became known as media contrast agents in clinical
protocols for Magnetic Resonance Imaging (MRI) and Magnetic Fluid
Hyperthermia (MFH). For these applications, Magnetite (\MAG) and
maghemite ($\gamma$-Fe$_{2}$O$_{3}$) nanoparticles are the preferred
phases because of their low toxicity, high saturation magnetic
moment at room temperature ($M_{S}$ $\sim$ $75$ and $82$ emu/g,
respectively) and the highest ordering temperatures among spinel
ferrites.\cite{TC} To improve stability and biocompatibility the
particles are coated with a polysaccharide and suspended in
water-based solvents. Despite the success attained in obtaining
efficient dispersions of iron oxide particles for MRI protocols, the
understanding of the effects of these coatings on the efficacy in
MRI or MFH applications (from which a material with desired
properties can be designed for specific purposes) is far from
complete. \cite{MRI0,MRI1,MRI2,MRI3}

Regarding the fundamental mechanisms that govern the magnetic
behavior of a magnetic colloid, the connection between
single-particle properties and collective behavior of the
\emph{ensemble} of magnetic nanoparticles has many subtle facets.
When entering the few-nanometer scale the magnetic behavior of a
single particle gradually differs from the corresponding bulk
materials, making increasingly inaccurate the description of the
collective behavior in terms of `bulk-like' entities. Dipolar
magnetic interactions between ferromagnetic nanoparticles are known
to affect the magnetic dynamics of the system through changes on the
average anisotropy energy barriers, which in turn determines the
relaxation times, magnetic hardness, and ordering temperature.
Indeed, the effects of interparticle interactions are not restricted
only to the anisotropy barriers but have also been proposed as
stabilizers of the magnetic order at both particle core and surface
\cite{SKU03}. In spite of sustained efforts to solve this problem
along the last decade, the exact mechanisms linking these properties
are still being discussed \cite{WEN01,LUI03}. Theoretical approaches
usually start from the original model proposed by Stoner and
Wolfarth \cite{STO48} for non-interacting, monodispersed
single-domain particles, and add some specific perturbation to the
collective or single-particle properties within different
mathematical landscapes. In this way microscopic mechanisms like
spin disorder, surface contributions and collective behavior for
strongly correlated particles have been added to the original
model.\cite{3,IGL04,5} On the other side, suitable experimental
systems for testing those models are still a key problem to be
solved, because of the difficulty of synthesizing samples with
different (controlled) particle size distribution and particle
interactions.\cite{6,ENN04,HAR03}

Magnetite is a mixed (Fe$^{2+}$ and Fe$^{3+}$) iron oxide with
spinel structure, composed by a cubic close packed oxygen array,
plus six ($2+4$) interpenetrating f.c.c. lattices of two
nonequivalent cation sites with fourfold ($8$a) and sixfold ($16$d)
oxygen coordination, labeled as A and B sites, respectively. The
cubic unit cell has lattice parameter $a$ $=$ $8.39$
$\textrm{{\AA}}$. \cite{LEO04}Magnetic interactions within A and B
sublattices are of ferromagnetic type, whereas the strong
antiferromagnetic A-B coupling yields a high Curie temperature
$T_{C}$ $=$ $850$ $K$. Bulk \MAG is magnetically soft, i.e., the
magnetization can be fully saturated in fields $H \lesssim$ $ 1 kOe$
for any crystal direction. \cite{KAK89} At room temperature, the
combination of first- and second-order magnetocrystalline anisotropy
constants $K_{1}$ $=$ $-13\times10^{4}$ erg/cm$^{3}$ and $K_{2}$ $=$
$-3\times10^{4}$ erg/cm$^{3}$ makes the easy axis of magnetization
to be along the crystallographic $[111]$ direction.\cite{ARA92} At
low temperatures magnetite undergoes a first-order transition,
namely the Verwey transition, located at $T_{V}$ $\sim$ $115-124$
$K$ depending on sample conditions.\cite{VER39} The most conspicuous
evidence of this transition is the sharp increase of the resistivity
by two orders of magnitude,\cite{VER39,REVIEWZIESE} but also other
properties such as magnetization, thermopower and reflectivity show
abrupt changes. Regarding the structural transition reported at
$T_{V}$, there is some consensus about the ortorhombic symmetry of
the low-T phase, but no conclusive structural model has yet
emerged.\cite {IIZ83,WRI02} Despite the different crystal structures
proposed for the low temperature phase, magnetic measurements have
shown that below $T_V$ the magnetocrystalline anisotropy energy can
be described as having orthorhombic symmetry with the easy
\textit{c} axis tilted $\sim 0.2^{ \circ}$ from the $<$1 0 0$>$
cubic axis, and the new \textit{a} (hard) and \textit{b} axes lying
on the former cubic $(0 0 1)$ plane.\cite{WAL02} The resulting
anisotropy is then uniaxial with constants $K_{a}$ $\approx
2\times10^{6}$ erg/cm$^{3}$ and $K_{b}$$=$ $21\times10^{4}$
erg/cm$^{3}$, considerably larger than the $K_1$ constant of the
high-T phase.\cite{ABE76,PAL63}

In this work we performed a detailed magnetic and structural
characterization of a colloidal suspension of dextran-coated
magnetite nanoparticles, with the purpose of systematically
characterize both the single particle properties and the resulting
collective behavior. The present paper is organized as follows: in
Sec. \ref{sec:2}, the  experimental procedures are described. In
Sec. \ref{sec:3} the results are presented and discussed in four
parts: Sec. \ref{sec:A} deals with the analysis of the structural
data. In Sec. \ref{sec:B} the results from magnetization
measurements are shown. Sec. \ref{sec:C} describes the M\"ossbauer
data obtained with different applied fields and temperatures.
Finally, the discussion and conclusions drawn from the present work
are given in Sec. \ref{sec:4}.

\section{\label{sec:2}EXPERIMENTAL PROCEDURE}

The samples studied in this work consisted of Dextran-coated
magnetite nanoparticles dispersed in a water-based ferrofluid
(ENDOREM\textmd{$^{TM}$}, Guerbet) used in clinical protocols as a
contrast agent for magnetic resonance imaging of the liver and the
spleen.\cite{GUERBET} The dispersion of nanoparticles was studied
\emph{as supplied}, i.e. suspended in an aqueous liquid carrier,
conditioned in sealed cylindrical sample holders ($5$ $mm$
diameter $\times$ $6$ $mm$ high). A second fraction of the sample
was lyophilized ($@$ $T=85$  $K$ and $P=10^{-4}$  $Torr$)
 to increase the particle concentration keeping the
particle size distribution unaltered. These \emph{as supplied} and
lyophilized samples will be labeled hereafter as END1 and ENDS,
respectively. Characterization through X-ray diffraction (XRD) was
performed in a Philips PW 1820 diffractometer using Cu$-K_{\alpha}$
radiation ($\lambda = 1.5418$ {\AA}) and Ni filter. Transmission
Electron Microscopy (TEM) images were performed using a $200$ $kV$
Model CM200 Philips electron microscope, conditioning the samples by
dropping an alcohol-powder suspension on a carbon-coated nickel
grid. The elemental analysis of Carbon, Hydrogen, Nitrogen and
Sulphur contents was performed with a Perkin-Elmer $2400$ CHNS
microanalizer. The M\"ossbauer spectroscopy (MS) measurements were
performed between $4.2$ and $296$ K in a liquid He flow cryostat,
with a conventional constant-acceleration spectrometer in
transmission geometry using a $^{57}$Co/Rh source with c.a. $50$
$mCi$ activity. For in-field M\"ossbauer measurements, the powder
samples were prepared between acrylic discs, and mounted in the bore
of a 140-kOe superconducting magnet, in a vertical
source-sample-detector setup such that the direction of gamma ray
propagation was parallel to the magnetic-field axis. For this setup,
a sine-shaped velocity waveform was used to minimize mechanical
noise. The spectra were fitted to Lorentzian line shapes using a
non-linear least-squares program, calibrating the velocity scale
with a foil of $\alpha$-Fe at $296$ $K$. When necessary, a
distribution of hyperfine magnetic fields, isomer shift and
quadrupole splitting have been used to fit the spectra.
Magnetization and ac magnetic susceptibility measurements were
performed in a commercial SQUID magnetometer both in
zero-field-cooling (ZFC) and field-cooling (FC) modes, between $1.8$
K $<$ T $<$ $250$ K and under applied fields up to H $=70$ kOe. The
frequency dependence of both in-phase $\chi'(T)$ and out-of-phase
$\chi''(T)$ components of the ac magnetic susceptibility were
measured by using excitation fields of $1 - 4$ Oe and driving
frequencies $0.01$ Hz $\leq$ $f$ $\leq$ $1500$ Hz.

\section{\label{sec:3}EXPERIMENTAL RESULTS}

\subsection{\label{sec:A}Structural Analysis}

%\begin{figure}
%\includegraphics [width=7.9cm]{NP1.eps}
%\caption{\label{fig:tem}} %Powder x-ray-diffraction profile of the
%lyophilized sample ENDS indexed with the (hkl) reflections of the
%cubic \MAG phase. The arrows correspond to organic materials. Upper
%left panel: selected-area from TEM images of \MAG nanoparticles
%where some surface faceting is observable.Upper right panel:
%histogram of the particle size populations as observed from TEM
%images. The solid line is the best fit using a log-normal
%distribution with $\left<d\right>_{TEM}$ $=$ $8.3\pm0.8$ $nm$ and
%$\sigma_d$ $=$ $0.6$.}
%\end{figure}

The XRD profile of the lyophilized powder (fig. \NP1) was composed
of broad lines that could be indexed with a cubic spinel crystal
structure. The presence of minor amounts of unknown organic
ingredients (used for coating and stabilizing the ferrofluid) were
also detected. A rough estimation of the average grain size
$\left<d\right>_{RX} = 7\pm1$ nm was obtained by applying the
Scherrer formula to the most intense XRD line of the spinel
structure, without including effects from crystal stress. This is in
agreement with previously reported low-temperature M\"ossbauer data
\cite{YOMMM,YOJAP} that showed the existence of \MAG as the only
phase, without evidences of the related $\gamma$-Fe$_{2}$O$_{3}$
(maghemite) or $\alpha$-Fe$_{2}$O$_{3}$ (hematite) phases.

%\begin{figure}
%\includegraphics [width=7.9cm]{NP2.eps}
%\caption{\label{fig:MTFC7T}}% Magnetization curve of END1 sample
%after field-cooling ($H_{FC} = 70$ kOe) from 300 K, and measured for
%increasing T values at the same field. The arrow indicates the
%melting point of the frozen suspension.}
%\end{figure}

Fig. \NP1 also shows the particle size histogram obtained by
measuring the diameter of about $200$ particles in Transmission
Electron Microscopy (TEM) images. The resulting histogram could be
fitted with a log-normal distribution, with median diameter
$\left<d\right>_{TEM} = 8.3 \pm0.8$ $nm$ and distribution width
$\sigma_d = 0.6$. It can be noticed from comparison between the
fitted curve and the experimental histogram that the counted number
of particles with $d$ $<$ $3$ $nm$ is considerably larger than the
expected from a pure log-normal size distribution with the fitted
parameters. The magnified TEM image of specific areas showed
particles with spherical-like shapes, although for few particles
with the largest diameters we observed faceted particles, as
selected in the inset of Fig. \NP1. The latter shape is probably
related to the high crystallinity of the particles reflecting the
cubic crystal habit of the \MAG phase.

\subsection{\label{sec:B}Magnetization Data}

Some of the basic magnetic features of the present particles
established previously \cite{YOMMM,YOJAP} are summarized here for
clarity.\cite{YONOTE} The particles were superparamagnetic (SPM) at
room temperature, and magnetization curves $M(T)$ taken in FC and
ZFC modes showed blocking temperatures $T_{B}$ of ca. $50 K$ (for $H
= 100$ Oe). \cite{YOMMM} Additionally, ac susceptibility data
$\chi(T,f)$ confirmed \cite{YOJAP} the thermally activated
(Arrhenius) nature of the blocking process, and indicated that the
effect of interparticle interactions was to increase the energy
barriers.

In order to measure the evolution of the saturation magnetization of
the particles with temperature, we applied a field of $H = 70$ kOe
at $T$ $=$ $300$ $K$ (i.e., above the melting temperature of the
liquid carrier) to the \textit{as provided} sample, cooled down the
system and measured the $M_{FC}(T)$ curve for increasing
temperatures keeping the same field. It can be seen from fig. \NP2
that these fully aligned particles show a nearly linear dependence
of the $M_{FC}(T)$ data, varying within $\approx13-7$ emu/g in the
$1.8 \leq T \leq 300$ K temperature range. All magnetization values
were converted into emu/g \MAG units by subtracting the mass of
organic components as determined from elemental CHNS analysis,
calibrated within $0.1$ $\%$ precision (Table \ref{tab:CHNS}). These
results were in agreement with the expected nanoparticle
concentration from the nominal composition of the ferrofluid (0.07
mol \MAG /l) and yielded saturation values $M_{S}=$ 10-12 emu/g,
much smaller than the expected 85-95 emu/g for bulk magnetite.
Additionally, the $M(T)$ curves did not show any indication of the
jump in the magnetization expected for the Verwey
transition.\cite{VER39} To explore possible aging (oxidation)
effects we compared our initial measurements with other runs
performed after several months, taken in three different
magnetometers, finding that the magnetic moment was reproduced
within few percent of precision. Additional support for the observed
reduction of the magnetic moment of the particles came from the
$M(H)$ curves both below and above the blocked temperature $T_{B}$
$\approx$ $40$ $K$ (see below). As will be discussed in Sec.
\ref{sec:C}, M\"ossbauer data showed no indication of other
iron-oxides of smaller magnetic moment, as could be expected in case
of particle oxidation (hematite Fe$_{2}$O$_{3}$) or hydration
(ferrihydrite). Therefore we attempted to verify these rather low
values by independent measurements, namely, the magnetic behavior in
the SPM state.

%\begin{figure}
%\includegraphics [width=8.0cm]{NP3.eps}
%\caption{\label{fig:fitmh}}% Fit of $M$($H$,250 K) curve for sample
%END1 using equation \ref{eq:loglan}. The values of $\overline{\mu}$
%and $\sigma_{\mu}$ obtained for the best fit are $290\mu_{B}$ and
%$1.68$, respectively.}
%\end{figure}

The magnetization $M_{SPM}(H,T)$ of an ensemble of non-interacting,
single-domain particles in the SPM state, each with magnetic moment
$\mu$, can be represented by a Langevin function $L(x)$ of argument
$x=\mu H/k_{B}T$. Therefore all data $M(H,T)$ should scale into a
single curve when plotted against $H/T$, provided that $T_B \ll T
\ll T_C$. Based on the log-normal distribution of particle sizes
obtained from TEM data, we modify the model above by using a
distribution-weighted sum of Langevin functions\cite{PRB_Ferrari}

\begin{eqnarray}
M_{SPM}=N\int_{0}^{\infty}\mu L\left(\frac{\mu
H}{k_{B}T}\right)f(\mu)d\mu,\label{eq:loglan}
\end{eqnarray}
where $f(\mu)$ is the log-normal distribution function
\begin{eqnarray}
f(\mu)=\frac{1}{\sqrt{2\pi}\sigma_{\mu}\mu}\exp\left(-\frac{\ln^{2}(\mu/\left<\mu\right>)}{2\sigma_{\mu}^{2}}\right)\label{eq:eq3}
\end{eqnarray}
of magnetic moments $\mu$. In this equation $\sigma_{\mu}$ is the
distribution width and $\left<\mu\right>$ is the median of the
distribution related to the mean magnetic moment $\mu_{m}$ by
$\mu_{m} = \left<\mu\right>
\exp(\sigma_{\mu}^{2}/2)$.\cite{PRB_Ferrari}

The best fit for sample END1 (for which dipolar interactions are
expected to be less important) yielded a median magnetic moment
$\left<\mu\right>$ $=$ $290$ $\mu_{B}$ and a distribution width
$\sigma_{\mu}$ $=$ $1.68$ for the highest measured temperature (see
figure \NP3). The fittings of $M(H)$ curves were performed for $T =
100, 150, 200$ and $250$ K. We have observed that the resulting
median value $\left<\mu\right>$ increased c.a $10 \%$ from 250 K to
100 K, which is likely to be related to the temperature dependence
of the magnetic moment $M(T)$ discussed above (fig. \NP2). The
distribution width $\sigma_{\mu}$ decreased $\approx 6 \%$ within
the same temperature range. To relate this moment distribution to a
particle size distribution we have assumed spherical geometry for
the particles, as observed from TEM data (i.e., $\left<\mu\right> =
M_{S}\frac{\pi}{6}\left<d\right>^{3}$) and the \emph{experimental}
saturation magnetization value $M_{S} = 7.2$ $emu/g$ from the $M(H)$
data at the same temperature. We have obtained $\left<d\right>_{MAG}
= 7.7\pm0.5$ $nm$ with  $\sigma_{d} = 0.56$. The median particle
size $\left<d\right>_{MAG}$ obtained with the measured $M_S$ value
is in close agreement with the $\left<d\right>_{TEM}= 8.3\pm0.8$
$nm$ from TEM micrographs, and the corresponding distribution widths
are identical within experimental accuracy. The outsized reduction
in the magnetic moment of the particles observed in M(T) curves
(fig. \NP2) in the blocked state is therefore supported,
independently, by the measured $M(H,T)$ curves in the SPM state. The
origin of this reduction will be discussed in Sec. \ref{sec:4}. We
note here that the latter procedure for obtaining the particle
magnetic moment $\overline{\mu}$ \emph{does not depend} on the
absolute magnetization values but on the ${\mu} H$ product for a
given temperature.

%\begin{figure}
%\includegraphics [width=8.0cm] {NP4.eps}
%\caption{\label{fig:HC}}% Temperature dependence of the coercive
%field $H_{C}(T)$ extracted from $M(T,H)$ cycles. The solid lines
%represents the linear fits using eq. \ref{eq:HCvsT} for samples ENDS
%and END1. The inset shows typical magnetization curves $M(H,T)$ for
%lyophilized sample ENDS taken at different temperatures. Note that
%$M(H)$ approaches to saturation for fields $H \gtrsim 10$ kOe.}
%\end{figure}

We have further characterized the transition from blocked to SPM
state by measuring the hysteresis curves $M(H)$ at different
temperatures between $1.8$ and $100$ $K$ for samples END1 and ENDS.
From the obtained $M(H,T)$ curves (fig. \NP4) it is observed that
the coercivity fields $H_{C}$ of both samples remain essentially
zero down to $T$ $\approx$ $17-18$ $K$, increasing appreciably below
this temperature. The temperature dependence of $H_{C}(T)$, as
extracted from $M(H,T)$ curves, is the expected for single-domain
particles, i.e.,

\begin{eqnarray}
H_{C}=H_{C0}\left[1-\left
(\frac{25k_{B}T}{K_{eff}V}\right)^{\frac{1}{2}}\right],\label{eq:HCvsT}
\end{eqnarray}
\noindent where $H_{C0} = 2\alpha K_{eff}/M_S$, and $\alpha = 0.48$
is a phenomenological constant used for a randomly oriented
distribution of easy axis.\cite{STO48}

Therefore, eq.\ref{eq:HCvsT} can be used for determining $K_{eff}$
from the slope of the $H_C(T)$ curves, through the product $E_a =
K_{eff}V$, using the extrapolated value $H_{C0}$. Expressing $E_{a}$
in terms of an anisotropy field $E_{a}=H_KM_{S}$, the effective
anisotropy constant $K_{eff}$ for a cubic system with easy axis
along the [111] direction is related to $K_{1}$ and $K_{2}$ through
$K_{eff}=\frac{4}{3}K_{1}+\frac{4}{9}K_{2}$, \cite{CUL72} and thus
$K_{eff}= 18.7\times10^{4}$ erg/cm$^{3}$ for bulk \MAG. The values
of $K_{eff}$ were extracted from the best fit of eq. \ref{eq:HCvsT}
for $T \leq 15$ K, and the calculation was made using \emph{both}
the mean value from TEM distribution $\left<d\right> = 8.3$ $nm$ and
the experimental $M_S = 12$ emu/g values from $M(H)$ data at low
temperatures. The resulting $K_{eff}$ values shown in Table
\ref{tab:KfromHc} are very close to the expected anisotropy of bulk
magnetite, suggesting that no major contributions other than
magnetocrystalline are important. These values are also in agreement
with our previous results obtained from low temperature M\"ossbauer
data \cite{YOJAP}. By extrapolating the linear fits to $H_{C}(T)$
$=$ $0$, we found values of $T_{B}$ $=$ $17.4$ $K$ and $18.5$ $K$
for samples END1 and ENDS, respectively. We mention that these
values obtained from the extrapolation of $H_{C}(T)$ are smaller
than the blocking temperatures obtained from ZFC curves, which we
associate to the different field regimes of each measurement (see
discussion in Sec. \ref{sec:4}). The extrapolated $ T \rightarrow 0$
$K$ values were $H_{C0}$ $=418$ Oe and $428$ Oe for samples END1 and
ENDS, respectively. It is worth noting from fig. \NP4 that the
saturation of the magnetization ($M$ $\geq$ $0.95$ $M_{S}$) is
attained for $H \approx 10$ kOe, the value of $M_{S}$ being
calculated from the $H^{-1} \rightarrow 0$ extrapolation in the M
$vs.$ $H^{-1}$ curves. These saturating fields are low when compared
with results from other nanoparticles of similar average sizes,
where non-saturating behavior extended for applied fields larger
than 70-90 kOe.\cite{NS1,NS2,NS3} For sample END1 the extrapolated
$M_S = 13.5$ emu/g value from the $M(H)$ curves at T $=1.8$ K (fig.
\NP4) is slightly larger than the $M_S = 12.9$ emu/g value from FC
data (fig. \NP2). The origin of this difference is probably related
to differences in particle alignment when the field is applied in
the liquid state (FC process) and below the blocking temperature
($M(H)$ curves).

\subsection{\label{sec:C}In-Field M\"ossbauer Data}

The reversal of the magnetic moment of single-domain particles is
governed by the rate of flipping over an energy barrier that is
proportional to the volume of the particle. This thermally activated
process is described by a  relaxation time $\tau$ that follows an
Arrhenius law \cite{BRO63}

\begin{eqnarray}
\tau=\tau_{0}\exp\left(\frac{E_a}{k_{B}T}\right),\label{eq:arrhenius1}
\end{eqnarray}

\noindent where $E_a$ is the energy barrier for the magnetic
reversion of the system, $k_B$ is the Boltzmann constant, and
$\tau_{0}$ is the pre-exponential factor (of the order of $10^{-9}$
$-$ $10^{-13}$ $s$).\cite{MOR90A,DOR97} At room temperature ($T \gg
T_{B}$) the relationship $\tau$ $<<$ $\tau_{L}$ (where $\tau_{L}$
$=$ $10^{-8}$ $-$ $10^{-9}$ $s$ is the Larmor precession time of the
nuclear magnetic moments\cite{MOR90A}) is verified, so the fast
relaxation makes the magnetic interactions to be averaged to zero
and thus yields a paramagnetic-like doublet. For temperatures below
$T_{B}$ (i.e., $\tau >> \tau_{L}$), the M\"ossbauer spectrum is
magnetically split. The hyperfine parameters at low temperature
reported in ref. \cite{YOJAP} were in agreement with other reports
on nanostructured magnetite \cite{GOY03,TAR03} and showed no
sizeable reduction of hyperfine fields, in contrast to other
nanoparticle systems \cite{BOD94,JIA99,YOCU,PUE01} where the
reduction effect was assigned to surface disorder effects. From eq.
\ref{eq:arrhenius1} it follows that the blocking temperature $T_{B}$
depends on the experimental window time $\tau_{M}$ of a given
measurement. It is well known that the ratio between the blocking
temperatures from different techniques like M\"ossbauer and
magnetization measurements (having typically $\tau_{M}$ $=$
$10^{-8}$ and $10^{2}$ s, respectively) inferred from eq.
\ref{eq:arrhenius1} can be as large as
$\frac{T_{B}^{moss}}{T_{B}^{mag}}\approx10$. The blocking
temperature for M\"ossbauer experiments are usually defined as the
temperature at which each of the blocked and SPM subspectra have
50$\%$ of the total spectral area. We estimated $T_{B}^{Moss}$ from
a series of M\"ossbauer spectra taken for $4.2$ $K$ $< T <$ $294$
$K$ (not shown) as being located somewhere between $110$ and $120$
$K$, which is about $2-3$ times the value obtained from
magnetization data.

%\begin{figure}
%\includegraphics [width=7.5cm]{NP5.eps}
%\caption{\label{fig:mossbH}}% \MOS spectra of sample ENDS at $4.2$
%$K$ at different fields $B_{app}$ applied along the $\gamma$-ray
%direction. Solid lines are the fitted experimental spectra (open
%circles), and dashed lines correspond to each component with the
%hyperfine parameters shown in Table \ref{tab:moss2}.}
%\end{figure}

In order to obtain local information of the rotational process of
the magnetic moments, we performed in-field \MOS measurements up to
$B_{app}= 120$ $kG$. In this kind of experiments the effective field
$B_{eff}$ actually observed is the vectorial sum of $B_{app}$ and
the internal hyperfine fields $B_{hyp}$,
$\mathbf{B_{eff}=B_{hyp}+B_{app}}$ thus providing useful information
about relative orientation between the local magnetic field at the
probe nucleus and the external field (i.e., between $B_{app}$ and
the magnetic sublattices). Figure \NP5 shows the resulting \MOS
spectra of sample ENDS at $4.2$ $K$ for $B_{app}$ $=$ $0$, $10$,
$30$, $50$, $80$, and $120$ $kG$, applied along the $\gamma$-ray
direction.  It can be seen from the figure that for increasing
applied fields two distinct spectral components are apparently
resolved. These two components originate from the ferrimagnetic
alignment of Fe atoms at A and B sites in the spinel structure, and
the assignment of each subspectra to A and B sites (see Table
\ref{tab:moss2}) is based on the zero-field spectrum.

It can be observed in figure \NP5 that the intensities of the second
and fifth lines change as the applied field increases, which is a
direct indication of the progressive reversal of the spins in the
direction of the field. The spectral intensities of each line of a
magnetically split spectrum are usually expressed as $3:x:1:1:x:3$,
where $x$ is the intensity of lines 2 or 5. For the present setup,
the values of $x$ span from $x=2$ for a randomly oriented system to
$x=0$ for full alignment in the direction of the external field.
More generally, the ratio of spectral areas ($A_{i}$) of lines 1 and
2 (or 5 and 6), can be related to the angle $\beta$ between the
magnetic quantization axis (assuming that the magnetic interaction
is dominant) and the direction of the $\gamma$-ray beam (identical
to the applied field $B_{app}$ in our experimental setup) through
the expression

\begin{eqnarray}
\sin^{2}\beta=\frac{6\Lambda}{4+3\Lambda},\label{eq:lambda}
\end{eqnarray}

\noindent where $\Lambda=A_{2}/A_{1}$ (or $A_{5}/A_{6}$). The
resulting values of $\Lambda$ and $\beta$ obtained from the fits at
different fields are displayed in Table \ref{tab:moss2}. For
$B_{app} = 0$ both subespectra could be fitted with the $\Lambda =
0.67$ value expected for a random powder. On the other side, we
observed different behavior for A and B sublattices under increasing
applied fields. Indeed, Table \ref{tab:moss2} shows that the
magnetic moments at A sites are already aligned for $B_{app} \geq
10$ $kG$, whereas the corresponding B-site moments show a gradual
rotation and remain slightly misaligned up to the largest field of
$B_{app} = 120$ $kG$. It is worthwhile to mention that the above
observation relies on the capability of separately fitting (i.e.,
experimentally resolve) lines 2 and 5 of each subspectrum. These
lines usually display strong overlap in ferrimagnetic materials,
yielding unreliable $\Lambda$ values from the fitting procedure. For
the present data, we were able to resolve clearly lines 2 and 5 from
each sublattice at $B_{app}= 120$ $kG$, using the obtained values of
intensities as the starting free parameters for fitting at lower
$B_{app}$ values.

It is known that ferrimagnetic nanoparticles display non-collinear
(canted) spin structure, although the location of these disordered
spins (i.e., at the surface, core, or both) is still being
discussed.\cite{CHI01, MOR03, SEP00} The actual spin structure is
known to be more complex than the expected from the original
Yafet-Kittel (YK) model of competing interactions, \cite{YAF52} and
neutron diffraction and \MOS studies have provided evidence that
spin canting can be either restricted to a single magnetic
sublattice or extended to both cation sites.
\cite{TRO00,GOYZN03,OLI99,CHI00}. The present in-field \MOS data
clearly show that magnetic moments at A-sites (and also a fraction
at B-sites) are magnetically 'soft', in agreement with the
saturating behavior observed in $M(H)$ (see fig. \NP4) at low
temperatures. Additionally, the resolution of the two magnetic
sublattices due to the application of a large external field makes
possible to identify a minority of magnetically hard spins as
located at B-sites in the spinel structure. Despite their very high
\emph{local} anisotropy, however, this small fraction does not seem
to contribute appreciably to the global magnetic anisotropy, as
observed from the magnetization data presented before.

\section{\label{sec:4}Discussion}

The reduction of magnetic moments in small particles ($1-10$ $nm$)
and thin films has been well documented along the last
years\cite{DOR97,BAT02} in both metallic and insulating
materials\cite{NS1,NS2,NS3,GOY03,JIA99}. A very-well accepted
explanation for the reduction of $M_{S}$ in ultrafine maghemite
particles was given by Berkowitz et al.\cite{BER68}, who proposed a
shell-core structure with a magnetically dead surface layer
originated from demagnetization of the (paramagnetic) surface spins.
If the above mechanism is assumed for the present particles, a
simple calculation using the bulk magnetization value
($M_{S}^{bulk}$ $=$ $90$ emu/g) would imply a core diameter
$\left<d\right>_{core}$ $=$ $2.5$ $nm$ and shell thickness $t$ $=$
$2.9$ $nm$. Such a thick paramagnetic surface layer, although not
strictly unphysical, appears very unlikely since M\"{o}ssbauer
spectra ruled out secondary Fe-containing phases, and CHNS
compositional data showed that the organic material is much less
than the necessary for such an organic coating. More recently, spin
canting has been proposed \cite{IGL04,KAC02,IGL01} as the mechanism
for the $M_S$ reduction in spinel nanoparticles, due to competing
interactions between sublattices that yield magnetic disorder at the
particle surface.\cite{KOD96} Experimental evidence for this model
has been found from M\"{o}ssbauer spectroscopy, and neutron
diffraction measurements. \cite{PUE01,JIA99,LIN95} Extremely low
values of $M_S$ have been measured in ferrimagnetic spinel particles
with $d \lesssim 15$ nm, attaining values as small as $\approx 0.06
M_S^{bulk}$, where $M_S^{bulk}$ are the corresponding values of the
bulk material. \cite{LUT98,MAR98,KOD96}. From the theoretical side,
Monte Carlo simulations using different models and
approximations\cite{KAC02,IGL01,TRO90,KAC00} have shown that the
reduction of $M_S$ is size-dependent, and originates in a highly
disordered spin surface with large magnetic anisotropy. The large
magnetic anisotropy of the surface could be, in turn, responsible
for the non-saturating $M(H)$ curves even at large ($H \gtrsim 100$
kOe) applied fields, in agreement with the observed behavior of many
spinel nanoparticles. \cite{YOSSC,YOCU,KOD99,MAR98}

For present \MAG particles, however, the reduced $M_S$ observed is
not accompanied by the linear increase in $M(H)$ at high fields
characteristic of canted systems. On the contrary, both the $M(H)$
curves and in-field M\"{o}ssbauer spectra at low temperature showed
that the magnetic saturation is attained at moderate ($H
\thickapprox 10$ kOe) fields. We are therefore induced to
hypothesize that, instead a thick spin-disordered surface layer, the
origin for the observed reduction in $M_S$ could be related to
internal compensation of the magnetic sublattices A and B in the
spinel structure. Further support to this idea comes from the
evolution of the relative intensities of the A and B subespectra for
increasing fields: it can be seen from Table \ref{tab:moss2} that
the intensity of the subespectrum assigned to B site increases at
expenses of the A subespectrum. The mechanisms that originate this
peculiar situation are not obvious, and further experiments should
be done before an explanation can be given.\\

Turning now to the measured magnetic anisotropy of the present
nanoparticles, we want to discuss the values extracted from
different techniques involving different experimental conditions. It
can be seen from Table \ref{tab:KfromHc} that the resulting values
of $E_a$ obtained from magnetization and \MOS data are essentially
coincident with the magnetocrystalline anisotropy of bulk
Fe$_{3}$O$_{4}$. Indeed, this is a remarkable result since shape
anisotropy is usually a major contribution to the average total
anisotropy \cite{CUENTA}. It is generally accepted that, when the
size of a magnetic particle decreases, surface effects become
increasingly important. However, symmetry arguments suggest that the
contribution from the surface should average to zero for a perfectly
spherical particle.\cite{BOD94} Moreover, numerical calculations
have shown\cite{GAR03} that for an unstrained simple cubic lattice
the bulk anisotropy vanishes in first-order, whereas the
contribution from the N\'{e}el surface anisotropy is second-order,
having the same cubic symmetry of the bulk. Therefore for highly
symmetric particle shapes such as cubic or spherical, the
contribution from both core and surface anisotropies should be
comparable to the bulk magnetocrystalline values. These symmetry
arguments are clearly applicable to our samples since, as mentioned
above, the low magnetic anisotropy values obtained for these
particles indicate magnetocrystalline anisotropy as the major
contribution, setting an upper limit (no more than $\sim$ $2$ $\%$)
for possible deviations from sphericity.\cite{CUENTA} Similar
arguments can be applied to the case of cubic-shaped particles like
the faceted ones observed in our TEM images, since demagnetizing
factors for a rectangular prism are within $15$ $\%$ of those of an
ellipsoid having the same $a$$=$$b$, $c$, and $r$$=$$c/a$
parameters, as calculated by Aharoni.\cite{AHA98} Therefore, the
above results support a picture of particles with spherical or
edge-rounded cubic shapes, having cubic crystal structure with
bulk-like anisotropy.

A third value for $E_{a}/k_B$ was previously \cite{YOJAP} estimated
from low temperature M\"{o}ssbauer data, using the model of
collective magnetic excitations (CME). \cite{MOR80} In this model,
small-angle fluctuations of the magnetic moments around the easy
magnetization axis, with characteristic times much shorter than the
\MOS sensing time $\tau_{L}$, yield a decrease of the observed
magnetic hyperfine field $B_{eff}(T)$ of \MOS spectra.\cite{MOR90A}
Expressing the magnetic energy $E$ $=$
$E(\alpha_{x},\alpha_{y},\alpha_{z})$ of a single particle as a
function of the direction cosines
$(\alpha_{x},\alpha_{y},\alpha_{z})$ of the saturation magnetization
$M_S$ relative to the crystallographic axes, the thermal dependence
of $B_{eff}$ derived for an arbitrary function of the energy can be
expressed as\cite{MOR83A}

\begin{eqnarray}
B_{eff}(T)=B_{0}\left(1-\frac{k_{B}T}{\kappa
V}\right),\label{eq:collective}
\end{eqnarray}

\noindent where $B_{0}$ is the hyperfine field at $T$ $=$ $0$,
and the effective anisotropy constant $\kappa$ is given by

\begin{eqnarray}
\frac{1}{\kappa}=1-\frac{1}{2}\left[\left(\frac{\partial^{2}
E}{\partial \alpha_{x}^{2}}\right)^{-1}+\left(\frac{\partial
^{2}E}{\partial\alpha_{y}^{2}}\right)^{-1}\right], \label{eq:eq10}
\end{eqnarray}

In a previous work on \MAG particles\cite{GOY03} with median sizes
$\left<d\right>$ ranging from $150$ to $5$ $nm$, it was reported
that for $\left<d\right>$ $\leq$ $50$ $nm$ no evidence of the Verwey
transition is observed, suggesting that those particles preserve the
cubic phase at low temperatures ($T < T_{V}$). Therefore, for the
present nanometric \MAG particles the values of magnetocrystalline
anisotropy constants $K_{1}$ and $K_{2}$ of the high-temperature
cubic phase with easy magnetization axis $[111]$ should be used at
any temperature. Applying these considerations and eq. \ref{eq:eq10}
an effective anisotropy constant
 $\kappa=-\frac{4}{3}(K_{1}+\frac{K_{2}}{3})$ is obtained. Using
 the $\left<d\right>_{TEM}$ value from the
lognormal fit to the TEM distribution, the best fit of $B_{eff}(T)$
gave an effective anisotropy $\kappa$ $=$ $24.3\times10^{4}$
erg/cm$^{3}$, in good agreement with the $18.7\times10^{4}$
erg/cm$^{3}$ expected using the bulk magnetocrystalline values
$K_{1}$ and $K_{2}$.

Inasmuch as the collective magnetic excitation model refers to the
thermal excitations of the lowest energy levels, for $T << T_{B}$,
the $E_{a}$ parameter obtained from eq \ref{eq:collective}, using
the CME model at zero field, gives information about single-particle
anisotropy properties since interparticle interactions will have
negligible effects on the detailed shape of the $E_{a}(\theta,H)$
surface in the vicinity and at the points of minima. On the other
side, the values extracted from the frequency-dependent maxima in
a.c. susceptibility data are related to the overcome of the magnetic
anisotropy when thermal energy equals the average height of the
energy barriers. The resulting $E_a$ values extracted from the
unblocking process can therefore be influenced by other
contributions (e.g. interparticle interactions) that affect the
values of the $E_{a}(\theta,H)$ surface near the
maxima.\cite{KOD99}\\

In order to experimentally check the influence of small applied
fields to the energy barriers, we measured the field-dependence of
$T_{B}$ through ac susceptibility curves $\chi(f,T,H)$ at different
applied dc fields. From the $\chi(f,T,H)$ data taken with $H = 10$
and $150$ Oe, we extracted the $T_{B}(f,H)$ dependences shown in
fig. \NP6. It can be observed that the frequency dependence
$T_{B}(\ln f)$ is linear for both applied fields, demonstrating that
the thermal activation mechanism of the blocking transition is
preserved. The best fits of $T_{B}^{-1}(\ln f)$ data were obtained
with $E_{a}/k_B = 436(9)$ and $296(15)$ K for $H = 10$ and $150$ Oe,
respectively, indicating that the effect of the applied field is to
decrease the effective anisotropy barrier $E_{a}(H)$. These results
give support to the idea that the application of small fields can
perturb the relative values and position of the maxima of the
anisotropy energy function.

%\begin{figure}
%\includegraphics [width=8.0cm] {NP6.eps}
%\caption{\label{fig:arrhenius}}% $T_{B}(\ln f,H)$ dependence
%extracted from $\chi(f,T,H)$ data taken with $H = 10$ and $150$ Oe.
%The solid lines represent the best fit with eq.
%\ref{eq:arrhenius1}.}
%\end{figure}

To further verify the above statements on a quantitative basis, we
calculated the energy curves for a single particle in applied fields
$H = 0$, $10$ and $150$ Oe by the following relation

\begin{eqnarray}
E(\theta,H)=K_{eff}V+\mu_{0}HM\cos\theta,\label{eq:eq9}
\end{eqnarray}

\noindent where $\theta$ is the angle between the easy magnetization
axis and the magnetization vector $M$ of the particle, and $H$ is
the external field, applied along the easy axis of magnetization. In
the presence of a small applied field $H$ parallel to the easy
direction of the particle (assumed here to be uniaxial), the
energies at the minima for $\theta = 0 ^{\circ}$ and $\theta = 180
^{\circ}$ will shift, and the transition probabilities between the
two minima will result in different relaxation times $\tau_{+}$ and
$\tau_{-}$ (i.e.,for forward and backwards relaxation processes,
respectively). In the present case of modest applied fields, the
observed relaxation times can be expressed as \cite{BRO63}

\begin{eqnarray}
\frac{1}{\tau}=\frac{1}{\tau_{-}}+\frac{1}{\tau_{+}},\label{eq:tau2}
\end{eqnarray}
\noindent with
\begin{eqnarray}
\tau_{\pm}= \tau_0
\exp\left({{E(\pm{H})/{k_BT}}}\right).\label{eq:tau1}
\end{eqnarray}

The energy curve for $H = 150$ Oe was computed using a gradient
method to calculate the local energy minima for each spin
configuration along the magnetic energy surface $E(\theta,H)$. Using
eqs. \ref{eq:eq9} - \ref{eq:tau1} and considering the
\emph{experimental values} $M_{S}$ $=$ $12$ emu/g,
$\left<d\right>_{TEM}$ $=$ $8.3$ $nm$ and $E_{a}$ $=$ $K_{eff}V$ $=$
$7.26\times10^{-14}$ $erg$ the resulting height for the energy
barrier ${\Delta E}/k_B = {E_{\theta=90}}/k_B - {E_{\theta=0}}/k_B$
was $500$ K at $T = 10$ K, in excellent agreement with the
experimental results in Table \ref{tab:KfromHc}. A systematic series
of computation for increasing applied fields $H$ from $0$ to $300$
Oe showed that the difference $\Delta E_{0}$ between the energy
maximum and the minimum at $\theta$ $=$ $0$ goes deeper whereas the
corresponding $\Delta E_{\pi}$ (minimum at $\theta$ $=$ $\pi$) gets
shallower. Therefore, the corresponding relaxation times $\tau_{+}$
and $\tau_{-}$ show opposite behavior for increasing fields $H$
provided that $\mu_{0}HM << K_{eff}V$. We also computed the ratio
$\tau_{+} / \tau_{-}$ for our experimental conditions. The results
obtained for $T = 10$ K and the experimental applied fields $H = 10$
and $150$ Oe (see fig. \NP6) yielded $\tau_{+} / \tau_{-} =
2\times10^{-1}$ and $7\times10^{-5}$, respectively. These results
give quantitative support to the assumption that the energy barriers
(i.e., the whole relaxation process) can be changed in the observed
amounts by the small fields used in our measurements. It should be
pointed out that an exact calculation of the relaxation time would
require, within Brown's model, to work with a probability function
of angles and time, which obeys a Fokker-Planck equation. For
symmetries different than uniaxial, the corresponding eigenvalues of
the Fokker-Planck equation have been obtained\cite{DOR97} by
numerical methods only for asymptotical situations
(i.e.,$\frac{k_{B}T}{KV}\rightarrow \infty$ or
$\frac{k_{B}T}{KV}\rightarrow 0$) not applicable to our present
case.

It was already mentioned that the values of $T_{B}$ obtained from
the extrapolation of $H_C(T)\rightarrow0$ from $M(H)$ curves at low
temperature differ from the ones obtained from ZFC measurements. The
difference between these two results is probably related to the
different field regimes in both techniques: while the ZFC maxima are
obtained at low fields (typically $100$ Oe), \emph{from a previously
demagnetized state}, the $H_{C}(T_B)$ is obtained after magnetic
saturation of the samples (we used $H_{max} = 30$ kOe). Therefore,
the moderate fields $H_C$ applied when magnetization reversal takes
place are likely to assist the unblocking process by decreasing the
energy barriers and the inferred $T_B$ will be lower than the
corresponding values extracted from the ZFC curves measured at lower
fields. The latter values, in turn, are mainly determined by the
dynamics of the unblocking process that depends on the local minima
at the energy surface originated from interparticle interactions.
The above interpretation is in agreement with our estimations of the
energy barriers $E_a / k_B$ from a.c. susceptibility measured at low
field,\cite{YOJAP} (reproduced in Table \ref{tab:KfromHc} for
clarity) which are larger than those extracted from the $H_C(T)$
fit.

During the analysis of the present data, the activation energies
values extracted from $\chi(f,T,H)$ data with $H = 0$ and $150$ Oe
were compared to the previously reported value\cite{YOMMM} of
$E_{a}$ $=$ $694$ $K$ for the same END1 sample, obtained with $H =
30$ Oe (see Table \ref{tab:KfromHc}). Contrary to the expected, the
value obtained at an intermediate dc field is \textit{larger} than
the two obtained at lower and higher fields. Having verified that
the temperatures of the maxima in our $\chi(f,T)$ curves were
reproducible within  $\Delta T$ $\leq $ $3$ $K$ on different
experimental runs, we attempted to investigate this seemingly
non-monotonic behavior of the anisotropy energy barriers by
measuring the dependence $T_{B}(H)$ from ZFC curves at different
fields. The results are shown in figure \NP7. It can be seen that
for sample END1 there is a continuous increasing of $T_{B}(H)$ for
decreasing fields, but a clear bending of $T_{B}$ towards smaller
values is observed for $H \leq 20$ Oe. The reproducibility of these
maxima in $M(T,H)$ curves was checked by performing a second set of
ZFC curves, after allowing the sample to demagnetize at room
temperature for several weeks. The results of the second run,
plotted in fig. \NP7 (black dots) demonstrate that the main features
of the $T_{B}(H)$ curve obtained on the second run remain unaltered.
This `reentrant` behavior has been originally found in a rather
different context, i.e., in $CuMn$ and $GdAl$ spin glasses, and
interpreted as due to the existence of a finite transition
temperature at zero field. \cite{BAR83} Theoretical evidence of
reentrant behavior of the Gabay-Tolouse line has been also found
using a classical $m-$vector model with uniaxial anisotropy.
\cite{VIE00}

%\begin{figure}[t]
%\includegraphics [width=8.0cm] {NP7.eps}
%\caption{\label{fig:TBvsH}}% The dependence $T_{B}(H)$ from ZFC
%curves of samples ENDS and END1 ($1^{st}$ and $2^{nd}$ run) at
%different fields.}
%\end{figure}

It seems more plausible, however, to relate this non-monotonic field
dependence of $T_B(H)$ at low fields with changes in the anisotropy
energy surface $E(\theta,H)$. It can be seen from fig. \NP7 that the
$T_{B}(H)$ curves tend to merge each other in the large ($\geq 500$
Oe) region, in agreement with the erasing of local minima by the
external field, and both curves steadily diverge for small applied
fields. However, for the concentrated sample a monotonic behavior
persists down to the smallest applied fields. This behavior has been
already found in diluted magnetite-based ferrofluids with different
concentrations,\cite{LUO91}, but no explanation for the
non-monotonic $T_{B}(H)$ dependence was offered. In a detailed work,
Sappey \textit{et al.} \cite{SAP97} proposed that the applied field
can produce a small broadening of the energy barrier distribution,
which suffices to give the observed field dependence of $T_B$. For
the present data, different behavior is observed in two samples with
identical particle size distributions but different average particle
distances. A rough estimation using the compositional data from
Table \ref{tab:CHNS} and $\left<d\right>_{TEM}=8.3$ $nm$, yielded
values of the average particle-particle distance $\left<e\right> =
63$ nm and $\left<e\right> = 2$ nm for END1 and ENDS, respectively.
Therefore in sample END1 we have $\left<e\right> \approx
8\left<d\right>$ and dipolar interactions are likely to be much
weaker than for the $\left<e\right>$ $\approx$ $\left<d\right> \over
4$ situation of concentrated ENDS sample. The above arguments
support the idea that the observed differences in $T_{B}(H)$ curves
are related to the influence of interparticle interactions on the
energy barrier. The absence of nonmonotonic behavior in the
concentrated sample indicates that the local dipolar field must be
included when considering the influence of the applied field on the
barrier distribution, as proposed by Sappey \textit{et
al.}\cite{SAP97}. Whatever the actual mechanism of the observed
behavior of $T_B(H)$, the values of $E_{a}$ in Table
\ref{tab:KfromHc} obtained at $H = 10$, $30$ and $150$ Oe can be
understood as having the same origin.

In conclusion, our present results on nanometer-sized magnetite
nanoparticles showed some intriguing facets. Single-particle
anisotropy energy both in blocked and SPM regimes and different
experimental time windows, indicates a reduced magnetization with
minimum spin disorder. The magnetic response of the system can be
described as `bulk-like` (regarding magnetic softness and
anisotropy) and, together with the reduced $M_S$ and the magnetic
saturation in moderate fields experimentally observed, can be
explained if a change in each sublattice magnetization is assumed.
This in turn might be originated from departures from the bulk
populations of the $Fe^{2+}$ and $Fe^{3+}$ ions at A and B sites
yielding a more complex magnetic configuration than the
two-sublattice structure. At low fields, particle interactions
become increasingly relevant, and may erase the non-monotonic
dependence of the blocking process observed in diluted samples. As
the energy barrier height and distribution usually determine the
dynamics of magnetic nanoparticle systems at low fields, its complex
dependence with external parameters has to be
solved before accurate parameters can be extracted from magnetic measurements.\\

\begin{acknowledgments}
We are indebted to Prof. R.F. Jardim for his suggestion of exploring
the field-dependence of the a.c. susceptibility, and Prof. A.
Labarta for suggestions and critical reading of the manuscript.
Thanks to Dr. A.B. Garcia for the compositional analysis. ELJ
acknowledges financial support from the VolkswagenStiftung, Germany,
through a Postdoctoral Fellowship. This work was supported in part
by Brazilian agencies FAPESP and CNPq.

\end{acknowledgments}

\pagebreak
\bibliography{apssamp}

\begin {references}
\bibitem{YONOTE} An error in ref. \cite{YOMMM} was brought into our
attention: The $M(H)$ cycle shown in the inset of fig. 2 of that
reference corresponds to FFMAG particles obtained from a sol-gel
route, not to the Endorem particles as stated in the caption.
\bibitem{CUENTA} For instance, deviations from ideal spherical shape
into a prolate spheroid with $r$ $=$ $c/a$ ($a$ $=$ $b$ and $c$ are
the minor and major axis, respectively) will add to the effective
anisotropy through a shape-anisotropy constant given by $K_{S}=\pi
M^{2}[1-\frac{3}{B^{2}}(\frac{r\ln(r+B)}{B}-1)]$.

where $M$ is the particle magnetization and $B=\sqrt{r^{2}-1}$.
Since shape anisotropy is proportional to $M^{2}$, a ratio $r$ as
small as $1.3$ will contribute with $K_{shape}$ $\sim$
$14\times10^{4}$ erg/cm$^{3}$, which is about the same value of the
bulk magnetocrystalline anisotropy.
\end{references}

\pagebreak
\begin{table}[t]
\caption{\label{tab:CHNS}Sample composition obtained from CHNS
elemental analysis of Carbon, Hydrogen, Nitrogen and Sulphur
contents. The [NP] column is the calculated particle concentration
(in particle/cm$^3$).}
\begin{ruledtabular}
\begin{tabular}{lccccc}
Sample & H (\%) &   C (\%) & N (\%) & S (\%) & [NP]  \\
\hline \vspace{-0.2cm}\\
\vspace{0.2cm}
END1 & 60.3(5)  & 26.4(5) & 3.3(5) & 1.1(5)  & $6.5\times10^{15}$\\
ENDS & 2.0(5)   & 24.2(5) & 0.8(5) & 0.0(5)  & $1.6\times10^{17}$
\end{tabular}
\end{ruledtabular}
\end{table}
TABLE I

\newpage
\begin{table}
\caption{\label{tab:KfromHc} Median particle diameter
$\left<d\right>$, effective anisotropy constant $K_{eff}$ and energy
barrier $E_a/k_B$ obtained from thermal dependence of coercive field
($H_{C}(T)$); Langevin fits of the SPM magnetization ($M_{SPM}$);
thermal dependence of the hyperfine fields
(eq.\ref{eq:collective}).}
\begin{ruledtabular}
\begin{tabular}{lccccccc}
&\multicolumn{2}{c}{$\left<d\right> (nm)$\footnotemark[1]} &
\multicolumn{2}{c}{$K_{eff}$ ($10^{4}$ erg/cm$^{3}$)
\footnotemark[2]}
&\multicolumn{2}{c} {$E_a/k_B$ (K)}& \\
&END1&ENDS&END1&ENDS&END1&ENDS&\\ \hline
$H_{C}(T)$  & 8.6 & 9.0 & 20.7 & 23.5 & 449 & 509 \\
$M_{SPM}$    & 7.7 \footnotemark[3]  & $7.9$\footnotemark[3] & $--$& $--$& $323$ \footnotemark[1]& $349$ \footnotemark[1]\\
Moss  & -- & 9.1 & -- & 24.3 & --   & 526 \\
$\chi_{ac}(f,T)$\footnotemark[4]& 9.9  & 10.3 & 32.1 & 35.1  & 694  & 763 \\
\end{tabular}
\end{ruledtabular}
\footnotetext[1]{Values calculated using $K_{eff}^{bulk} = 18.7
\times10^{4}$ erg/cm$^{3}$. } \footnotetext[2]{Values calculated
using $\left<d\right>_{TEM} = 8.3$ nm.} \footnotetext[3]{Calculated
using the experimental $M_S = 7.2-7.3$ emu/g values at $T = 250 K$.}
\footnotetext[4]{From Ref.\cite{YOJAP}.}
\end{table}
TABLE II

\begin{turnpage}
\begin{table}
\caption{\label{tab:moss2} Hyperfine parameters at different applied
fields $B_{app}$. $B_{eff}$ is the effective field, $I_{T}$ is the
total intensity of each six-line subespectrum, $\Lambda$ is the
ratio of intensities of lines 2 and 1 $(A_{2}/A_{1})$, and $\beta$
is the canting angle of each sublattice.}
\begin{ruledtabular}
\begin{tabular}{lcccccccccccc}
$B_{app}$ (kG) &\multicolumn{2}{c}{0}&\multicolumn{2}{c}{10
}&\multicolumn{2}{c}{30 }&\multicolumn{2}{c}{50
}&\multicolumn{2}{c}{80 }&\multicolumn{2}{c}{120}\\ \hline
Sites&A&B&A&B&A&B&A&B&A&B&A&B\\ \hline \vspace{-0.2cm}\\

$B_{eff}$ (kG) & 526(1) & 508(1) & 530(1) &  505(1)  & 540(1) &
496(1) & 565(1)  &  479(1) & 592(1) & 449(1)
    & 632(1) & 411(1) \\
$IS$ (mm/s)& 0.18(1) & 0.24(1)& 0.17(1)& 0.24(1)& 0.15(1)& 0.23(1)
&0.18(1)& 0.26(1)& 0.13(1) &0.25(1)&
0.13(1) &0.25(1) \\
$QS$ (mm/s)& 0.03(2)& -0.02(2)& 0.03(2)& -0.03(2) &0.00(2)&
-0.04(1)&
0.01(1)& -0.04(2) &0.02(1)& -0.01(2)& -0.01(1)& -0.01(1)  \\
$\Gamma$ (mm/s) & 0.63(3)& 0.86(4) &0.45(4) &0.79(4) &0.69(4)
&0.86(3)& 0.62(4) &0.83(4)
&0.70(4)& 0.82(3) &0.75(4)& 0.84(4) \\
$I_{T}$ (\%) & 35(2)& 65(2) &34(3)& 66(2)& 35(3)& 65(2) &38(2)&
62(3)& 60(2)& 40(3) &61(2)& 39(3) \\
$\Lambda$& 0.70(3) &0.68(3)& 0.04(3)&0.23(3)&0.00(4)& 0.16(4)&
0.02(3)& 0.13(3)&0.00(3)&0.09(3)& 0.00(3) & 0.10(3)\\
$\beta$\footnote{\hspace{0.15cm} Calculated from eq.\ref{eq:lambda}}
(degrees)
               &&&0(4)&34(4)&0(4)&31(4)& 0(4)& 31(4)& 0(4)& 22(4)& 0(4)&22(4) \\
\end{tabular}
\end{ruledtabular}
\end{table}
\end{turnpage}

\newpage
\section*{FIGURE CAPTIONS}

Figure \NP1: Powder x-ray-diffraction profile of the lyophilized
sample ENDS indexed with the (hkl) reflections of the cubic \MAG
phase. The arrows correspond to organic materials. Upper left panel:
selected-area from TEM images of \MAG nanoparticles where some
surface faceting is observable.Upper right panel: histogram of the
particle size populations as observed from TEM images. The solid
line is the best fit using a log-normal distribution with
$\left<d\right>_{TEM}$ $=$ $8.3\pm0.8$ $nm$ and
$\sigma_d$ $=$ $0.6$.\\

Figure \NP2: Magnetization curve of END1 sample after field-cooling
($H_{FC} = 70$ kOe) from 300 K, and measured for increasing T values
at the same field. The arrow indicates the
melting point of the frozen suspension.\\

Figure \NP3: Fit of $M$($H$,250 K) curve for sample END1 using
equation \ref{eq:loglan}. The values of $\overline{\mu}$ and
$\sigma_{\mu}$ obtained for the best fit are $290\mu_{B}$ and
$1.68$, respectively.\\

Figure \NP4: Temperature dependence of the coercive field $H_{C}(T)$
extracted from $M(T,H)$ cycles. The solid lines represents the
linear fits using eq. \ref{eq:HCvsT} for samples ENDS and END1. The
inset shows typical magnetization curves $M(H,T)$ for lyophilized
sample ENDS taken at different temperatures. Note that
$M(H)$ approaches to saturation for fields $H \gtrsim 10$ kOe.\\

Figure \NP5: \MOS spectra of sample ENDS at $4.2$ $K$ at different
fields $B_{app}$ applied along the $\gamma$-ray direction. Solid
lines are the fitted experimental spectra (open circles), and dashed
lines correspond to each component with the hyperfine
parameters shown in Table \ref{tab:moss2}.\\

Figure \NP6: $T_{B}(\ln f,H)$ dependence extracted from
$\chi(f,T,H)$ data taken with $H = 10$ and $150$ Oe. The solid
lines represent the best fit with eq. \ref{eq:arrhenius1}.\\

Figure \NP7: The dependence $T_{B}(H)$ from ZFC curves of samples
ENDS and END1 ($1^{st}$ and $2^{nd}$ run) at different
fields.\\

\end{document}